\documentclass[aps,prl,superscriptaddress,twocolumn]{revtex4-2}

\usepackage{float}
\usepackage{graphicx}
\usepackage{color,soul}

\begin{document}

\title{Universal electronic synthesis by microresonator-soliton photomixing}


\author{Jizhao Zang}
\email[]{jizhao.zang@nist.gov}
\affiliation{Time and Frequency Division, National Institute of Standards and Technology, Boulder, Colorado, USA}
\affiliation{Department of Physics, University of Colorado, Boulder, Colorado, USA}

\author{Travis C. Briles}
\affiliation{Time and Frequency Division, National Institute of Standards and Technology, Boulder, Colorado, USA}

\author{Jesse S. Morgan}
\affiliation{Department of Electrical $\&$ Computer Engineering, University of Virginia, Charlottesville, Virginia, USA}

\author{Andreas Beling}
\affiliation{Department of Electrical $\&$ Computer Engineering, University of Virginia, Charlottesville, Virginia, USA}

\author{Scott B. Papp}
\email[]{scott.papp@nist.gov}
\affiliation{Time and Frequency Division, National Institute of Standards and Technology, Boulder, Colorado, USA}
\affiliation{Department of Physics, University of Colorado, Boulder, Colorado, USA}



\begin{abstract}
Access to electrical signals across the millimeter-wave (mmW) and terahertz (THz) bands offers breakthroughs for high-performance applications. Despite generations of revolutionary development, integrated electronics are challenging to operate beyond 100 GHz. Therefore, new technologies that generate wideband and tunable electronic signals would advance wireless communication, high-resolution imaging and scanning, spectroscopy, and network formation. Photonic approaches have been demonstrated for electronic signal generation, but at the cost of increased size and power consumption. Here, we describe a chip-scale, universal mmW frequency synthesizer, which uses integrated nonlinear photonics and high-speed photodetection to exploit the nearly limitless bandwidth of light. We use a photonic-integrated circuit to generate dual, microresonator-soliton frequency combs whose interferogram is fundamentally composed of harmonic signals spanning the mmW and THz bands. By phase coherence of the dual comb, we precisely stabilize and synthesize the interferogram to generate any output frequency from DC to $>$1000 GHz. Across this entire range, the synthesizer exhibits exceptional absolute fractional frequency accuracy and precision, characterized by an Allan deviation of $3\times10^{-12}$ in 1 s measurements. We use a modified uni-traveling-carrier (MUTC) photodiode with an operating frequency range to 500 GHz to convert the interferogram to an electrical signal, generating continuously tunable tones across the entire mmW band. The synthesizer phase noise at a reference frequency of 150 GHz is -83 dBc/Hz at 100 kHz offset, which exceeds the intrinsic performance of state-of-the-art CMOS electronics. Our work harnesses the coherence, bandwidth, and integration of photonics to universally extend the frequency range of current, advanced-node CMOS microwave electronics to the mmW and THz bands.

\end{abstract}
\maketitle

Phase-stable frequency synthesis is a principal technology for universally important applications in communication, information processing, ranging, positioning, timing, and sensing \cite{Jones2000,Cundiff2001,taraF2011,Daryl2018,Tara2019}. Digital electronics make possible completely phase-coherent signal generation and detection across the microwave band, enabling for example the data stream central to mobile devices with increasingly rich functionalities. However, the growing demand for information capacity in wired and wireless networks \cite{Koenig2013} and emerging sensing approaches with high resolution \cite{Lien2016,Pauli2017} highlight the need for access to higher frequency bands, which are generally more challenging to implement with digital electronics.

The mmW and THz bands extend from 30 to 300 GHz and to tens of THz, respectively. Beyond increased bandwidth, resolution, and speed for applications, the mmW and THz offer unique opportunities to characterize materials \cite{Beard2002,Jepsen2011}. Hence, there exists a long history of development for mmW and THz sources, such as free-electron vacuum devices \cite{Booske2011}, solid-state diodes \cite{Midford1979,Eisele2004} , quantum cascade lasers \cite{Faist1994}, and integrated electronic circuits \cite{Chun-Cheng2012}. By building on miniaturized voltage-controlled oscillators, phase-locked loops, electronic filters, frequency multipliers, and other silicon microelectronic components, digital CMOS frequency synthesizers \cite{Musa2011, Chun-Cheng2012,Luo2013,Kim2019,Meng2019} define the state-of-the-art for widespread applications in the mmW and THz bands, including signal analysis, radar sensing, and high-speed communications. However, the phase noise of electronic synthesizers is fundamentally limited by frequency division and multiplication.

Photonic techniques for mmW synthesis generally involve laser sources that are heterodyne-detected to generate an electronic signal. Numerous photonic technologies have been reported to implement mmW synthesizers, including frequency multiplexing with electro-optic (EO) modulators \cite{Lin2009}, mixing two lasers locked to optical frequency combs \cite{Fukushima2003,Musha2004},  and Brillouin lasers using fiber cavity or disk resonators \cite{JiangLi2013,YihanLi2019}. The approaches that use modulators are challenging or impossible to scale to high-frequency signals due to the low power of high-order modulation sidebands. Millimeter-wave generation using fiber frequency combs provides a wide frequency-tuning range, but such systems cannot be integrated and they suffer from the low efficiency of the comb source. Synthesizers with Brillouin lasers feature high frequency stability, but the frequency tuning range is very narrow because both the pumps and Stokes light need to match the resonance of the cavity. Until now, a compact frequency synthesizer with tuning across the mmW and THz bands has not been demonstrated.

\begin{figure*}[ht!] \centering \includegraphics[width=1\textwidth]{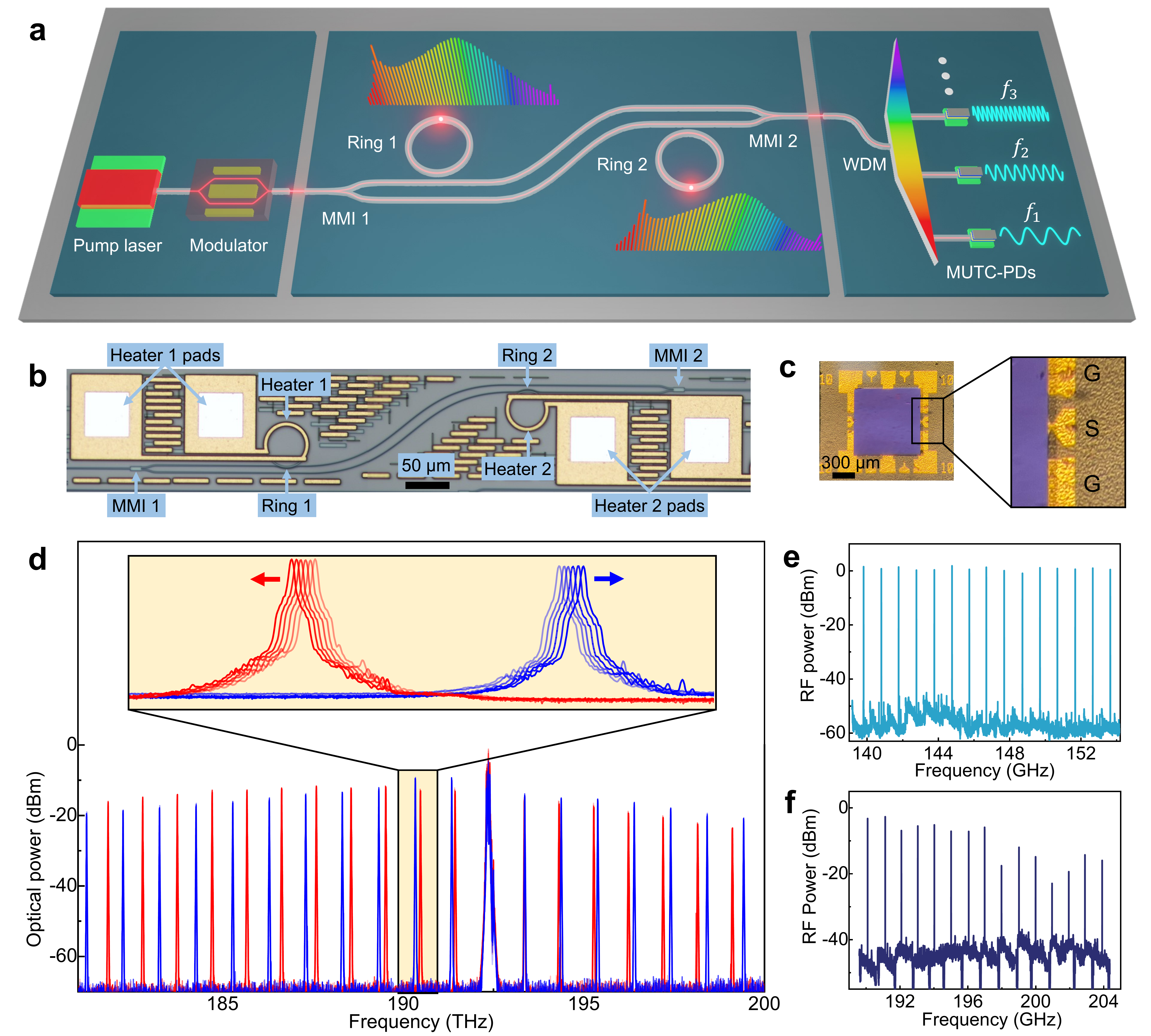}
\caption{A universal mmW frequency synthesizer architecture using soliton microcomb photomixing. a, Conceptual diagram of the chip-scale mmW synthesizer; b, Integrated soliton dual-comb generator; c, Microscope image of the high-speed modified uni-traveling-carrier photodiode, which is flip-chip bonded on the aluminium nitride sub-mount with ground-signal-ground coplanar waveguide; d, Comb spectra of $f_{\text{IM}}$ tuning, for example the inset shows an enlarged view of the mode pairs with $\approx150$ GHz spacing; e and f, Electrical spectra of the 150 GHz (e) and 200 GHz (f) beat notes with a frequency tuning step size of 1 GHz. The vertical axis represents the power of intermediate frequency signal after frequency down-conversion and power amplification.}
\end{figure*}

The recent development of Kerr-microresonator solitons \cite{TobiasHerr2014,Yi2015,Xue2015,Kippenberg2018,Su-peng2021,Sun2023} has provided a promising laser source to realize chip-scale electronic frequency synthesizers, leveraging photonic integration with lasers, modulators, photodetectors, and other nonlinear optical elements. Such soliton microcombs intrinsically operate with repetition frequency in the microwave and mmW bands, and photodetecting them yields a low phase noise electronic signal \cite{junqiu2020,xuyi2021,junqiu2025}. Based on development for Tb/s optical data links, microcombs can offer high output power and high conversion efficiency \cite{Victor2023,QiFan2024,Zang2025}. Single microcomb schemes for electronic signal generation can offer high performance spectral characteristics \cite{Papp2014,Newman2019,Erwan2020,Igor2024,Sun2024}, but the frequency tuning range is narrow and the signal stability is still limited by thermal noise \cite{Jordan2018} without the use of complex phase-locking with respect to optical frequency references.

Here, we introduce a universal, electronic frequency synthesizer by way of microresonator-soliton photomixing. The synthesizer is a photonic-integrated circuit (PIC) with parallel-operating soliton microcombs and singular input and output ports for the pump laser and the dual microcomb, respectively. By heterodyne detection of the two soliton microcombs, we generate a set of highly coherent photocurrent tones that continuously span the microwave, mmW, and even THz frequency ranges.  Intensity modulation of the pump laser provides a frequency bias of the dual microcomb that continuously and completely tunes the synthesizer output. Indeed, we achieve mmW synthesis at 100 GHz, 150 GHz, to a maximum of 300 GHz with an absolute fractional frequency accuracy of  $3\times10^{-12}$, limited in our experiments by the reference clock. The phase noise of the synthesizer at 150 GHz is -83 dBc/Hz at 100 kHz offset, which outperforms the state-of-the-art frequency synthesizers in 65 nm CMOS electronics \cite{Musa2011,Luo2013,Amir2018,Meng2019}. Leveraging the advantages of soliton microcombs such as photonic integration, low noise, and low power consumption, our work demonstrates an architecture for a chip-scale frequency synthesizer that generates stable electrical signals from DC to $>$1000 GHz with better agility and phase noise than frequency realized in advanced node CMOS.

Figure 1 presents an overview of the electronic synthesizer, showing the concept of microresonator-soliton photomixing with integrated photonics and the synthesizer's exceptional frequency-tuning range. We highlight the key photonic elements of the synthesizer in Fig. 1a, including the dual-microresonator-soliton generator, the pump laser, the intensity modulator that creates two pump tones with a tunable frequency offset $\Delta \nu_{p}=2f_{\text{IM}}$, where $f_{\text{IM}}$ is the modulation frequency for each microcomb, and the wavelength filtering and photodetection system that maps each mode pair of the dual microcomb to an electronic frequency. In our experiments, we explore different versions of the synthesizer with partial photonic integration (Fig. 1b) and with discrete microcomb chips, a bench laser system, and fiber-integrated components. The key operational principle of the synthesizer is to employ two soliton microcombs with a difference in repetition frequency of $\Delta\nu_{\text{rep}}$ and a difference in pump frequency of $2f_{\text{IM}}$, hence commensurate mode pairs of the microcombs have an increasing and tunable frequency spacing of  $|\Delta\nu_{\text{rep}} \pm 2f_{\text{IM}}|$, $|2\Delta\nu_{\text{rep}} \pm 2f_{\text{IM}}|$, $|3\Delta\nu_{\text{rep}} \pm 2f_{\text{IM}}|$,  and so on. 

Fully integrating the synthesizer is an important goal. While no platform yet exists to realize heterogeneous integration of high-performance lasers, intensity modulators, Kerr microresonators, and photodetectors, future platforms are likely to support these technologies. Our advance of partial integration-- the dual soliton microcomb system --highlights the potential for combination with other technologies; see Fig. 1b. Indeed, integrated photonics with silicon or InP waveguides has been heavily developed, making optical modulation, wavelength division, and connections to high-speed electronics available \cite{Liu2010,Dong2016,Marko2018}. We have developed wafer-scale fabrication of soliton microcombs by several iterations of fabrication and testing with the Ligentec foundry \cite{Zang2024}. For integration of the soliton system, we employ a silicon nitride (Si$_3$N$_4$, hereafter SiN) waveguide layer with bottom and top SiO$_2$ claddings to implement the microcombs, which are joined to common chip-edge couplers with multimode interference (MMI) couplers.  The ring resonance is tuned by the on-chip thermal heaters with a tuning rate of ~0.52 nm/V at 1550 nm. By optimizing the bias voltage applied to the heaters, we are able to align the ring resonances with the two pump tones, respectively.   

To convert mode pairs of the dual-soliton microcomb to an electronic signal, we use a MUTC photodiode \cite{Qinglong2016,Jesse2018,Zang2025mixer}; see Fig. 1c. This photodiode structure is specially designed for high-frequency and high-power operation by only allowing fast electron transition through the depleted region. The overall photodiode bandwidth depends on the transit time and $RC$-limited bandwidth. The former refers to the time delay between carrier generation in the absorption layers and carrier collection in the contacts, while the later is mainly induced by the photodiode's series resistance and junction capacitance.  We use an absorption layer of 150 nm and drift layer of 350 nm to reduce the transit time-limited bandwidth, and the photodiode diameter is 5 $\mu$m, reducing the junction capacitance for large RC-limited bandwidth. The photodiode is flip-chip bonded to an aluminium-nitride submount to efficiently remove heat and simplify mmW measurement with a ground-signal-ground (GSG) probe. The coplanar waveguide (CPW) on the submount is optimized to extend the overall bandwidth by inductive peaking. Using this design, we realize a photodiode with 3 dB bandwidth of 128 GHz, responsivity of 0.17 A/W, output power over -8.6 dBm at 160 GHz \cite{Jesse2018}, and operational response to 500 GHz \cite{Zang2025mixer}.

\begin{figure*}[ht!] \centering \includegraphics[width=1\textwidth]{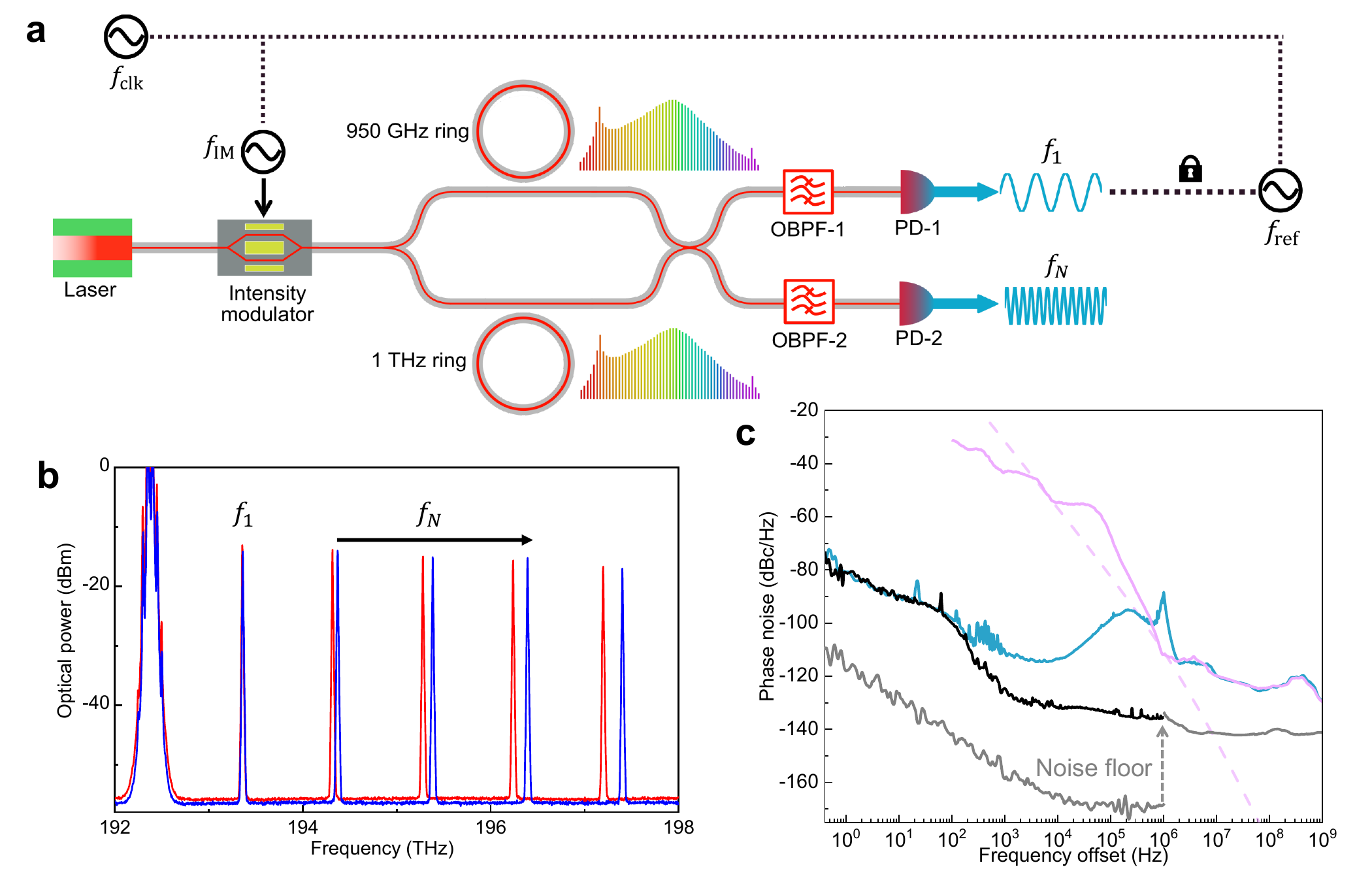}
\caption{Millimeter-wave frequency synthesis by locking $f_{1}$ to a stable reference clock $f_{\text{ref}}$. a, Experimental setup, including a CW laser, an intensity modulator driven by $f_{\text{IM}}$, two microresonators on separated chips, two optical band pass filters, a low-speed photodiode (PD-1) to detect $f_1$ and a high-speed MUTC photodiode (PD-2) to detect $f_N$;  b, Optical spectrum of the dual soliton combs, emphasizing the first mode pair ($f_1$) for locking $\Delta\nu_{\text{rep}}$ and high-order mode pairs ($f_N$, $N>1$) for mmW generation; c, A comparison of $f_1$ phase noise between the free-running (solid pink trace) and phase-locked (blue trace) cases. The contribution of $f_{\text{ref}}$ and thermal noise is shown by the black and dashed pink traces, respectively. }
\end{figure*}

By tuning $f_{\text{IM}}$, we synthesize mmW signals with an extremely wide tuning range. Figure 1d presents optical-spectrum measurements of the synthesizer output with varying $f_{\text{IM}}$. We use two soliton microcombs with 1 THz and 950 GHz spacing, namely $\Delta\nu_{\text{rep}}$=50 GHz, and tune $f_{\text{IM}}$  around 25 GHz. The frequency spacing  between the commensurate mode pairs is an integer multiple of 50 GHz like 100 GHz, 150 GHz, 200 GHz, and so on. The upper inset is an enlarged view of the optical spectra, where we take the second mode pair in the lower frequency side of the pump as an example. We generate mmW signals by detecting the mode pairs with the MUTC photodiode. Figure 1e and f show the electrical spectra of 150 GHz and 200 GHz mmW signals, respectively, with horizontal axis representing the corresponding power after frequency down conversion. We tune the mmW frequency across a range of 13 GHz with a step size of 1 GHz. In Fig. 1f, we observe a fast roll-off of IF power as frequency increases, which results from the increased insertion loss of the GSG waveguide probe at frequencies beyond its nominal operational bandwidth. Our approach based on $f_{\text{IM}}$ tuning allows for simultaneous tuning of the spacing  between each mode pair. With a $f_{\text{IM}}$ range of 12.5 GHz,  our electronic synthesizer is able to seamlessly cover the whole mmW band and higher frequencies.

\begin{figure*}[ht!] \centering \includegraphics[width=1\textwidth]{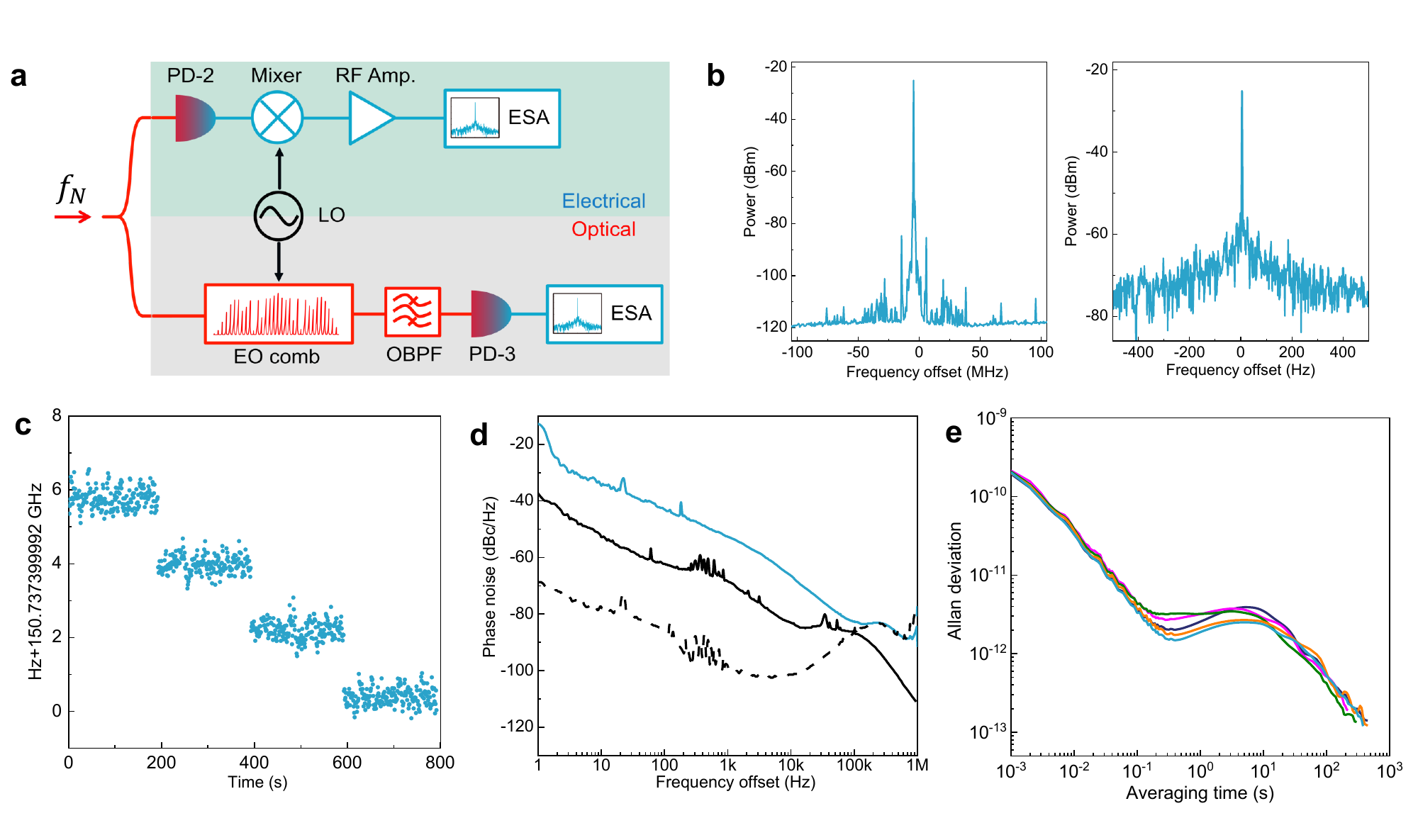}
\caption{Characterization and fine frequency tuning of the synthesized mmW signals. a, Two methods of implementing frequency down-conversion: using an even harmonic mixer (upper panel) and electro-optic combs (lower panel); b, Electrical spectra of the 151 GHz mmW signal with a frequency span of 200 MHz (left) and 1 kHz (right), respectively; c, We finely tune 150 GHz mmW with a step size of 1.8 Hz and wait for 200 seconds at each frequency; d, Measured phase noise of  $f_4$ (blue trace). For comparison, we plot the phase noise contributed from the term $6f_{\text{IM}}-10f_{\text{LO}}$ (solid black trace) and  the term $4f_1$ (dashed black trace); e, Measured Allan deviation of $f_N$ for N=3,4,5,6,7 (orange: $f_3=100$ GHz, blue: $f_4=150$ GHz, dark blue: $f_5=200$ GHz, magenta: $f_6=250$ GHz, green: $f_7=300$ GHz).}
\end{figure*}

Phase stabilizing $\Delta\nu_{\text{rep}}$ is the essential aspect of our electronic synthesizer. The key advantage is that the mmW frequency noise depends on $\Delta\nu_{\text{rep}}$ instead of  the repetition rate of each soliton microcomb, therefore a fully stabilized microcomb is not required. We leverage the common-mode pump laser and near-perfect equidistance of the separate microcombs to transfer the stability of a low-frequency microwave clock ($f_{\text{clk}}$) directly to the mmW and THz ranges. The synthesizer output comes directly from $\Delta\nu_{\text{rep}}$ and $f_{\text{IM}}$. In particular, the spectra of the two microcombs are 
\begin{equation}
    \nu_{\text{N1}}=\nu_{\text{p1}}+N \nu_{\text{rep1}}
    \label{EQ:950G comb line}
\end{equation}
and 
\begin{equation}
    \nu _{\text{N2}}=\nu _{\text{p2}}+N{\nu_{\text{rep2}}}
    \label{EQ:THz comb line}
\end{equation}where $\nu_{N}$, $\nu_{p}$ and $\nu_{\text{rep}}$ are the comb frequencies, the pump frequencies, and the repetition frequency, respectively. The integer $N$ is the comb mode index with respect to the pump, and it is negative (positive) for the comb modes at lower (higher) frequency with respect to the pump. Indeed, the optical heterodyne of commensurate mode numbers from each comb yields the fundamental output of our electronic synthesizer, namely
\begin{equation}
    f_{N}=\mid \nu_{\text{N1}}-\nu_{\text{N2}} \mid = \mid 2f_{\text{IM}}-N\Delta\nu_{\text{rep}}\mid
    \label{EQ:beat frequency}
\end{equation}
where $2f_{\text{IM}}=\nu _{\text{p1}}-\nu _{\text{p2}}$ and $f_{\text{IM}}$ is derived from $f_{\text{clk}}$, namely $f_{\text{IM}}=\alpha f_{\text{clk}}$, where $\alpha$ is a frequency multiplication factor. 

Figure 2 presents an overview of the phase-locking system and the fundamental phase-noise performance of the synthesizer, including a system diagram (Fig. 2a), an optical spectrum showing mode pairs for locking $\Delta\nu_{\text{rep}}$ ($f_1$) as well as mmW generation at $f_N$ (Fig. 2b), and single-sideband (SSB) phase-noise measurements (Fig. 2c). As shown in Fig. 2a, the pump laser is modulated by an intensity modulator, generating two sidebands that serve as the pumps for two SiN microresonators with free spectral range (FSR) of 950 GHz and 1 THz, respectively.  We initiate solitons in both microresonators simultaneously by fast sweeping the pump frequency across each resonance \cite{Jordan2018}. The two soliton microcombs are combined by a fiber coupler, followed by two tunable optical band pass filters (OBPFs) to select different mode pairs for photodetection in a low speed photodiode (PD-1) and a high-speed MUTC photodiode (PD-2), respectively. 

To synthesize the mmW output $f_{N}$,  we detect and lock the microcomb mode pair that corresponds to  $f_1= 2 f_{\text{IM}}-\Delta\nu_{\text{rep}}$ ( $2 f_{\text{IM}}>\Delta\nu_{\text{rep}}$ in our experiment). By applying electronic feedback to the pump frequency, we lock $f_1$ with a low-frequency reference $f_{\text{ref}}$ that is synthesized to $f_{\text{clk}}$, namely $f_1=f_{\text{ref}}=\beta f_{\text{clk}}$. Then $\Delta\nu_{\text{rep}}$ is expressed as: 
\begin{equation}
   \Delta\nu_{\text{rep}}=2f_{\text{IM}}-f_{1}=(2\alpha-\beta)f_{\text{clk}}
    \label{EQ:repetition rate difference}
\end{equation}
Substituting Equation \ref{EQ:repetition rate difference} into Equation \ref{EQ:beat frequency}, we can derive the expression for $f_{N}$ as a function of $f_{\text{clk}}$:
\begin{equation}
    f_{N}=\mid 2\alpha(1-N)+N\beta \mid f_{\text{clk}}
    \label{EQ: MMW frequency}
\end{equation}
where $\alpha$, $\beta$ and $N$ are integers. Equation \ref{EQ: MMW frequency} shows that $f_{N}$ is a phase-coherent multiplication of $f_{\text{clk}}$. After photodetection in the high-speed MUTC photodiode, we achieve precise frequency synthesis between a high-frequency $f_{N}$ and the low frequency $f_{\text{clk}}$, with user-defined integer ratios ($\alpha, \beta$) and mode index ($N$).

\begin{figure*}[ht!] \centering \includegraphics[width=1\textwidth]{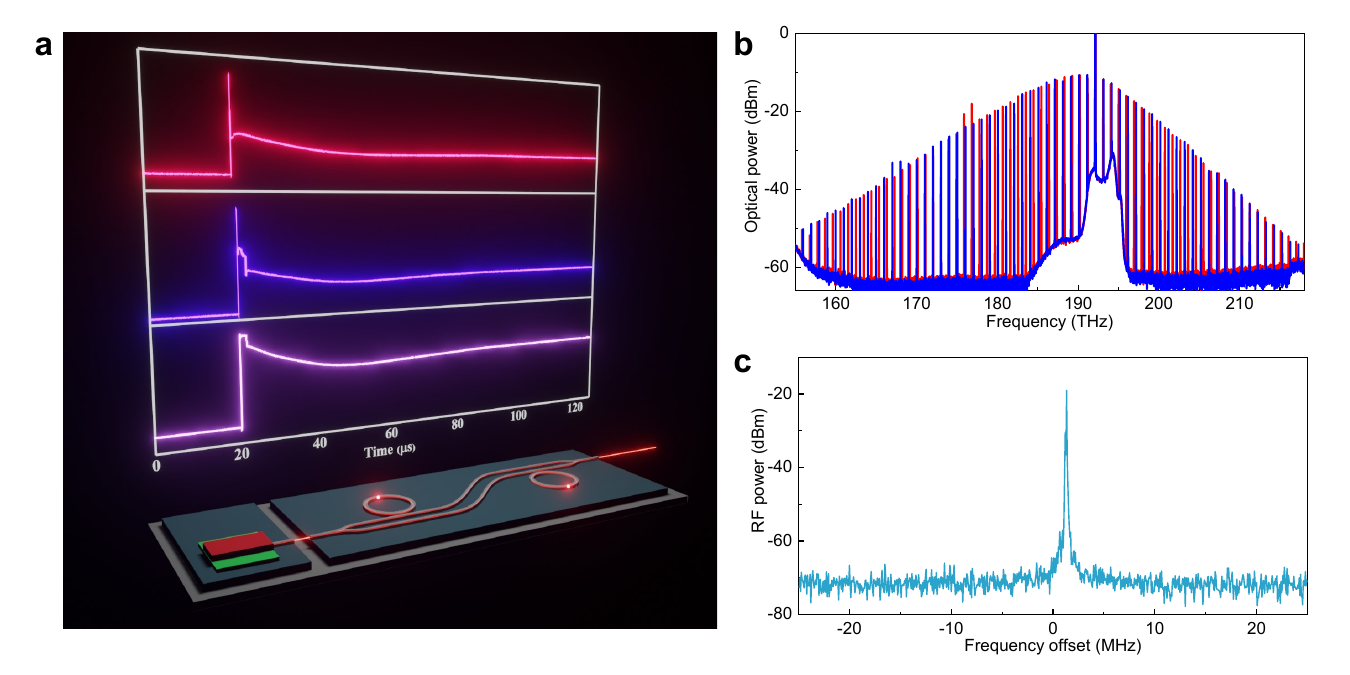}
\caption{Soliton microcomb generation and photomixing using the integrated dual-comb generator. a, oscilloscope traces of the comb power in the case with 956 GHz soliton only (upper panel), 1005 GHz soliton only (middle panel) and both simultaneously (lower panel); b, Optical spectra of the two soliton microcombs with $\nu_{\text{rep}}$ of 956 GHz (red) and 1005 GHz (blue); c, Electrical spectrum of the 147 GHz beat note corresponding to $f_{-3}$, which is measured with a resolution bandwidth (RBW) of 30 kHz.}
\end{figure*}

Phase-noise analysis reveals that $f_{1}$ noise is mainly determined by $f_{\text{ref}}$ noise, the PLL noise, and the pump laser frequency noise. Figure 2c presents a comparison of the phase noise between the case with free-running (solid pink trace) and locked $f_{1}$ (blue trace). The noise floor of the phase-noise analyzer is shown by the gray trace with discontinuity resulting from different phase noise analyzers that are used below and above 1 MHz offset frequency. For reference, we show the noise spectrum of $f_{\text{ref}}$ at 100 MHz (black trace) and the predicted soliton microcomb thermal-noise contribution (dashed pink trace), respectively. Below 400 Hz offset frequency, the phase noise of $f_{1}$ is primarily  determined by that of $f_{\text{ref}}$; from 400 Hz to 2 MHz, it is limited by the PLL noise; above 2 MHz, the frequency offset is beyond the PLL bandwidth, and the phase noise should follow the predicted thermal noise. However, the noise level does not decrease sharply as we would expect based on microcomb thermal noise. Instead, there is a noise floor of ~-120 dBc/Hz at frequency offset above 1 MHz. We attribute this excess noise floor to the frequency noise of the pump laser. Equations \ref{EQ:950G comb line} and \ref{EQ:THz comb line} suggest that the pump frequency noise is transferred to each comb line directly or indirectly via the dependence of repetition frequency on detuning \cite{Jordan2018}. The 3-dB fiber couplers before and after the dual rings form an Mach–Zehnder interferometer (MZI) structure. After passing through this fiber MZI and photodetection in PD-1, the pump frequency noise is further transferred to the beat note, resulting in the phase noise floor modified by the MZI transfer function. Photonic integration of the synthesizer will eliminate this effect.

Next, we investigate the phase-noise performance of $f_N$ for N=3, 4, 5, 6, and 7. Since $f_N$ is beyond the operational bandwidth of our electrical spectrum analyzer (ESA) and phase-noise analyzer, we use two approaches for frequency down-conversion to characterize $f_N$; see Fig. 3a. In the first, we use a tunable 0.8 nm OBPF to select different mode pairs for photodetection.  The beat notes at mmW frequencies are derived from the high-speed photodiode (PD-2) by a GSG waveguide probe at 110 GHz-170 GHz. An even harmonic mixer (EHM) working with a stable 15 GHz local oscillator (LO) down-converts the high-frequency beatnotes to IF signals at $\approx100$ MHz.  In the second, the frequency down-conversion is realized with two cascaded electro-optic modulators \cite{Tara2019}. Two comb modes are phase-modulated to generate sidebands that are spectrally overlapped. We use a narrow-band optical filter to select the two sidebands which are close but derived from different comb lines, and detect $f_{\text{IF}}$ with a low-speed photodiode (PD-3). In both cases, we convert $f_N$ to an electrically detectable $f_{\text{IF}}$ with the help of the stable $f_{\text{LO}}$, namely $f_{\text{IF}}=f_{N}-M f_{\text{LO}}$, where $M$ is a frequency multiplexing factor.  Since we use a stable $f_{\text{LO}}$ and its noise contribution is negligible, we can characterize $f_{N}$ indirectly by measuring the low-frequency $f_{\text{IF}}$. Both methods characterize our mmW synthesizer output $f_N$ of up to 300 GHz, and the results of the approaches are in agreement.

Benefiting from the intrinsic low noise of the soliton microcombs, our electronic synthesizer generates  mmW signals with a high signal-to-noise ratio (SNR). We experimentally generate up to 251 GHz mmW signal by photomixing mode pairs in the high-speed photodiode. For example, we present $f_4$ of 151 GHz in Fig. 3b , where the left and right panel show the electrical spectra with frequency span of 200 MHz and 1 kHz, respectively.  The corresponding $f_{\text{LO}}$ is 15.084 GHz and its 10th harmonic is used for frequency down-conversion in EHM, allowing for $f_{\text{IF}}$ below 1 GHz. The SNR of 151 GHz signal is 93 dB, with a few spurs in the electrical spectrum that are induced by PLL and $f_{\text{ref}}$. 

We demonstrate the frequency stability of our electronic synthesizer by its fine frequency tuning (Fig. 3c), phase noise (Fig. 3d), and Allan deviation (ADEV, Fig. 3e) performance. Figure 3c depicts a linear frequency sweep of the mmW signal at 150 GHz ($f_4$), which is measured by a frequency counter with a gate time of 1 second after frequency down-conversion. The step size of frequency sweep is 1.8 Hz and we dwell for 200 seconds at each frequency to accumulate statistics on the synthesizer drift and fluctuation. The well-defined frequencies at such small tuning step, combined with the coarse tuning results in Fig. 1e, highlight the exceptional frequency tuning performance of our mmW synthesizer. Figure 3d shows the measured SSB phase noise of $f_4$ (blue trace) in the case with EHM for frequency down-conversion, showing a phase noise of -83 dBc/Hz at 100 kHz offset. According to Equation \ref{EQ:repetition rate difference} and Equation \ref{EQ:beat frequency}, $f_{\text{IF}}$ can be expressed as $f_{\text{IF}}=f_4-10f_{\text{LO}}=6f_{\text{IM}}-4f_1-10f_{\text{LO}}$. For comparison, we plot the phase noise contributed from the term $6f_{\text{IM}}-10f_{\text{LO}}$ (solid black trace) and the term $4f_1$ (dashed black trace). Below 30 kHz frequency offset, the phase noise of $f_4$ is $\approx20$ dB higher than the predicted value, which we attribute to the noise induced by the 3-dB fiber coupler that combines the two soliton microcombs for photomixing \cite{Schneider2013,Wu2025}. Since the phase delay induced by the optical fiber is sensitive to ambient environmental factors, such as temperature and vibration, the phase difference of each mode pair fluctuates with time, leading to increased phase noise. This phase fluctuation also results as a floor in the measured ADEV traces; see Fig. 3e. Here we use the EO modulation approach to characterize the ADEV of $f_{N}$, since there is less bandwidth limitation comparing with the second approach, using the high-speed photodiode and EHM. Despite the ADEV floor at 0.3 s to 10 s measurement time, from 100 GHz to 300 GHz the synthesizer output always reaches an frequency instability below $3\times10^{-12}$ at 1 second.  More photonic integration of the MMI couplers and both soliton microcombs on one chip (Fig. 1b) will eliminate this ADEV floor and further improve the frequency stability.

As the first, major step towards on-chip system integration, we create a dual-microcomb generator by integrating the 950 GHz and 1 THz rings, thermal heaters, and MMI couplers on one chip. Figure 4a illustrates soliton generation with our integrated dual-microcomb, which is shown in the photograph of Fig. 1b. We tune the pump modes of the two resonators to the same frequency and use a CW laser as the pump for simplification. As with the bench-top system, we initiate the soliton by a fast pump-frequency sweep.  After filtering out the pump, we monitor the comb power by use of a low-speed photodiode and an oscilloscope.  The inset of Fig. 4a shows the oscilloscope traces as we initiate the soliton in 956 GHz ring (upper panel), 1005 GHz ring (middle panel) and both simultaneously (lower panel). The corresponding microcomb optical spectra are shown in Fig. 4b with  blue and red traces represent the 1005 GHz and 956 GHz soliton, respectively. For example, the comb line pairs corresponding to $f_{-3}$ are selected by a tunable OBPF, followed by heterodyne detection with the high-speed MUTC photodiode. Figure 4c presents the beat note at 147 GHz after frequency down-conversion to 946 MHz and spectral measurement with a RBW of 30 kHz. The successful generation of mmW signal demonstrates the potential of a fully integrated mmW synthesizer.

In summary, we have introduced a universal mmW frequency synthesizer, based on dual microresonator soliton photomixing. By pumping the two microresonators with a single intensity modulated laser and phase-locking the lowest order mode pair of the dual microcomb, we generate exceptionally stable and tunable mmW signals, which are phase-coherently synthesized with respect to an input reference clock. We characterize the mmW synthesizer output over the range of DC--300 GHz and demonstrate an ADEV below $3\times 10^{-12}$ in 1 s measurements, limited by the reference clock. Using a high-speed MUTC photodiode, we electrically generate 150 GHz mmW signal with phase noise of -83 dBc/Hz at 100 kHz offset. The exceptional frequency tunability of our mmW synthesizer is highlighted by the coarse and fine tuning of 150 GHz mmW with a step size of 1 GHz and 1.8 Hz, respectively. Finally, we design a partially integrated version of our system and demonstrate 147 GHz mmW generation by photomixing mode pairs from the integrated dual-comb generator. Our work highlights an architecture for universal electronic synthesis, paving the way toward low-power, chip-scale mmW and THz frequency synthesizers with exceptional frequency stability and tunability.

\section*{Data availability}

The data that support the findings of this study are available from the corresponding author on reasonable request.

\section*{Code availability}
The simulation codes used in this study are available from the corresponding author on reasonable request.

\section*{Acknowledgment}
We acknowledge Madison Woodson and Steven Estrella from Freedom Photonics for MUTC PD fabrication. We thank Alexa Carollo and Haixin Liu for technical review of the letter. This research has been funded by the AFOSR FA9550-20-1-0004 Project Number 19RT1019, DARPA DODOS, NSF Quantum Leap Challenge Institute Award OMA – 2016244, and NIST. This work is a contribution of the U.S. government and is not subject to copyright. Trade names provide information only and not an endorsement.

\section*{Disclosures}
S.B.P. and J.Z have filed an US patent (11336377) regarding the technology reported in this article. The authors declare no other competing interests. 

\section{References}
\bibliography{sample}


\section{Methods}

\textbf{Soliton generation in SiN microresonators}. Our SiN devices are fabricated by Ligentec. The device layer thickness is 780 nm and it is covered by  3 $\mu  m$-thick silicon oxide cladding. The intrinsic quality factor of  950 GHz and 1 THz rings is $3 \times10^{6}$. To generate the 950 GHz soliton, we use a microresonator with ring radius of 24.5 $\mu m$ and ring width of 1620 nm. In the case with 1 THz soliton, the two parameters are 23.3 $\mu m$ and 1600 nm, respectively. The comb formation process in microresonators include several comb states, which highly depends on the pump-resonance detuning, pump power and cavity dispersion. We access the stable soliton states using a rapid frequency sweeping technique. First  we optimize the laser frequency and modulation frequency $f_{IM}$ to align two modulation sidebands with two microresonators' resonances. Then the two sidebands are coupled into a dual parallel Mach–Zehnder modulator (DPMZM) whose modulation signal is derived from a VCO.  We operate the DPMZM at single-sideband suppressed-carrier modulation (SSB-SC) mode to induce a common pump frequency shift for both microresonators. This pump frequency shift is determined by the VCO and varies with VCO's bias voltage (tuning rate $\approx$ 1 GHz/V). By applying an appropriate ramp signal to the VCO, we are able to fast sweep the pump frequency and simultaneously excite solitons in both microresonators. 

 \textbf{Coarse frequency tuning of the mmW synthesizer}. Wideband mmW tuning is realized by $f_{IM}$ tuning, which equivalently varies the pump frequencies because the laser frequency is fixed. During this process, it is challenging to keep the solitons running in both microresonators  because they only exist in a certain detuning range. Since the resonant frequency highly depends on the device temperature, we develop a computer-controlled lookup-table method to thermally tune the resonant frequency so that it can track the pump frequency automatically and keep the detuning in the soliton existence range. In Fig. 1e and f, We are able to keep the soliton state in both microresonators during the whole tuning range of 13 GHz, which is limited by the operating bandwidth of our RF amplifier used for intensity modulation.

\end{document}